\documentclass[preprint,showpacs,preprintnumbers,amsmath,amssymb]{revtex4}

\usepackage{textcomp}
\usepackage{makeidx}
\usepackage{amsmath}
\usepackage{subfigure}
\usepackage{amssymb}
\usepackage{hyperref}
\usepackage{graphicx}
\usepackage{epstopdf}

\begin{document}

\title{Study of branes with variable tension.}

\author{Rodrigo Aros}
\email{raros@unab.cl}
\affiliation{Departamento de Ciencias Fisicas, Universidad Andres Bello, Av. Republica 252, Santiago,Chile}
\author{Milko Estrada}
\email{mestrada@udla.cl}
\affiliation{Instituto de Matem\'atica, F\'isica y Estad\'istica, Universidad de las Am\'ericas, Manuel Montt 948, Providencia,Santiago, Chile}

\date{\today}

\begin{abstract}
In this work we study a brane world model with variable tension which gives rise to four dimensional cosmologies. The brane worlds obtained corresponds to E\"{o}tv\"{o}s branes whose (internal) geometry can be casted as either a four dimensional (A)dS$_{4}$ or a standard radiation period cosmology. The matter dominated period is discussed as well.
\end{abstract}

\keywords{Randall Sundrum, brane world, de Sitter, hierarchy problem , brane tension}
\maketitle

\section{Introduction}
A largely open question in physics has been the huge difference between the values of the Higgs mass, $m_H \approx 1 TeV$, and the Planck mass $m_P \approx 10^{19}GeV$, also called the \emph{hierarchy problem}. In 1999 Lisa Randall and Raman Sundrum \cite{7} proposed a model of two statics 3-branes with constant tensions of equal magnitude but opposite signs imbedded in an AdS$_5$ space. Our observed universe corresponds in this model to the positive tension brane. The geometry considered in that model can be described by the metric
\begin{equation}
ds_5^2=e^{-2kr|\phi|}  \eta_{\mu \nu}dx^\mu dx^\nu + r^2 d\phi, \label{RS1}
\end{equation}
where $e^{-2kr|\phi|}$ is called the warp factor. The coordinate system $x^M = (x^{\mu},\phi)$, with $x^{\mu}=(t,x,y,z)$, satisfies $-\infty< x^{\mu} <\infty$  and $\phi \in [-\pi,\pi[$. Our universe is located at $\phi=0$ and the secondary brane, which is called the strong brane, at $\phi=\pi=-\pi$. The hierarchy problem is solved in this model due to the mass in the two branes is related by $m_{\phi=0}=e^{-kr\pi}m_{\phi=\pi}$. This allows the Planck mass at strong brane to be of same order of Higgs mass at our universe provided $e^{kr\pi} \approx 10^{15}$.

In RS model both branes have constant tension and the geometry is static. This last, however, does not describe an evolving universe. To do that the brane section should be promoted at least to (a flat-FRW)
\begin{equation}\label{Promotion}
  \eta_{\mu \nu}dx^\mu dx^\nu \rightarrow -dt^2 + a(t)^2 \mathbf{dx} \cdot \mathbf{dx}.
\end{equation}

From a geometrical point of view Eq.(\ref{Promotion}) is not a minor change. A solution is to consider a brane world scenario with variable tension branes. This implies that the branes considered cannot be fundamental branes of the standard fashion. However, this not all, and another aspect to be addressed is establishing the dependency on the temperature of the evolution of the brane-universe or viceversa. In Ref. \cite{17} was  proposed that the tension of our brane universe should depend on the temperature of the universe according to {\it  E\"{o}tv\"{o}s law}:
\begin{equation}
T=K(x_c-x), \label{2}
\end{equation}
where $K$ is a constant, $x$ represents the temperature and $x_c$ is a initial temperature of our universe. Motivated by Stefan-Boltzmann law, where the energy density of the CMB is proportional to $x^4$, it can be proposed that $x \propto a^{-1}$ and therefore \eqref{2} can be rewritten as:
\begin{equation}
T = K x_c \Big (1- \frac{a_{min}}{a(t)} \Big )  \label{3} ,
\end{equation}
where $a_{min}$ is the initial value of scale factor on our universe. This model is called  E\"{o}tv\"{o}s branes, see ref. \cite{17}, and predicts that as our universe expands, and the temperature decreases, the tension of our brane universe increases, becoming more and more rigid. This model is compatible with the current observations. It must be stressed that, at least in principle, this model does not constraint the strong brane dependency on time or temperature.

In \cite{20} Abdalla, da Silva and da Rocha proposed a modification of the E\"{o}tv\"{o}s branes where the brane tension becomes the linear function of $t$, the FRWL \emph{time}, \begin{equation}
T=\pm \lambda t  \pm  \beta \label{4},
\end{equation}
where $\lambda$ and $\beta$ are positive constants. The final result of the model is a universe that as it expands and cools the brane tension increases. This reproduces most of the features of the E\"{o}tv\"{o}s branes. Unfortunately this model does not provide a direct solution for the scale factor $a(t)$ in the FRWL metric.

In work we aim to test a toy model of variable tension branes that reproduce a E\"otv\"os brane for the universe and simultaneously constraints the form $a(t)$ by the five dimensional Einstein equations. To consider non constant tension branes requires to propose a model for the brane. Following  \cite{20}, and in order to simplify the computations, the only change, with respect to a fundamental brane, is to replace the constant tension by
\[
T+ \frac{t}{k} \frac{dT}{dt}
\]
where $k$ is a dimensionless parameter. The idea behind this is to introduce a conformal expansion along the FRW time. This choice allows to obtain different solution for $T$, using the Brane World Sum Rules of the same fashion as Eq.\eqref{3} but without the need to impose a priori an E\"{o}tv\"{o}s tension.

The number of solutions is large, but since this work is not aimed to represent the complete scenario of a variable tension brane world model, the analysis will be restricted to only three cases of physical interest. It is worth noticing that this model is compatible with those discussed in \cite{Brane-Localized} and references there in.

In the next section, we derive the Brane World Sum Rules method. Next we will find the tensions of both branes and the scale factor.

\section{ Brane World Sum Rules Method}
Brane World Sum Rules is a set of consistency conditions derived from the Einstein equations for brane world scenarios with spatially periodic internal space. See \cite{18}. This method allows to find relations between the cosmological constant $\Lambda_{5D}$, the Ricci scalar $R^{(4D)}$ and the tensions of both branes. Let's consider the line element:
\begin{equation}
ds^2= W^2(\phi)g_{\mu\nu}(x^\alpha)dx^\mu dx^\nu+r^2 d\phi^2,  \label{6}
\end{equation}
where $x^{\alpha}=(t,x,y,z)$ and $x^5=\phi$ takes the values $-\pi<\phi<\pi$. $r$ is \emph{radius} of compactification. This yields
\begin{align}
&R_{\mu\nu}^{(5D)}=R_{\mu\nu}^{(4D)}-\frac{3}{r^2} g_{\mu\nu}(x^\mu)(W')^2 -\frac{1}{r^2} g_{\mu\nu}(x^\mu) W W''  ,  \label{7} \\
&R_{55}^{(5D)}=-\frac{4W''}{W} ,  \label{8}
\end{align}
where $'$ indicates differentiation with respect to $\phi$. Multiplying the equation \eqref{7} by  $g^{\mu\nu}_{(5D)}=W^{-2}g^{\mu\nu}(x^\mu)$ and (\ref{8})  by $g^{55}=\frac{1}{r^2}$ determine that
\begin{align}
&R^{\mu (5D)}_\mu-R^{(4D)}W^{-2}=-\frac{12}{r^2}(W')^2W^{-2}-\frac{4}{r^2}W''W^{-1} ,    \label{9} \\
&R^{5(5D)}_5=-\frac{4}{r^2} W''W^{-1}.  \label{10}
\end{align}

In the same fashion, multiplying equation \eqref{9} by $(1-n)W^n$ and \eqref{10} by $(n-4)W^n$ and adding both equations, yields
\begin{align}
&\frac{R^{\mu(5D)}_\mu-R^{(4D)}W^{-2}}{12}(1-n)W^n+\frac{R^{5(5D)}_5(n-4)W^n}{12}= \nonumber \\
&\frac{W^n}{r^2} \Big (\frac{(n-1)(W')^2}{W^2}+\frac{W''}{W} \Big ).  \label{11}
\end{align}
Now, since
\begin{equation}
\frac{(W^n)''}{n}=W^n \Big (\frac{(n-1)(W')^2}{W^2}+\frac{W''}{W} \Big )   , \label{12}
\end{equation}
the relation \eqref{11} can be written as:
\begin{align}
\frac{(W^n)''}{r^2 n}&=\frac{R^{u(5D)}_u-R^{(4D)}W^{-2}}{12}(1-n)W^n \nonumber \\
                                   &+\frac{R^{5(5D)}_5(n-4)W^n}{12}.   \label{13}
\end{align}
By using the Einstein equation $\frac{R_{AB}}{8\pi G_{5}}=T_{AB}-\frac{g_{AB}}{3}T$, we write down the following equations:
\begin{align}
&  R^{\mu(5D)}_\mu= \frac{ 8\pi G_5 }{3} \Big  (-T^\mu_\mu- 4 T^5_5 \Big ) ,         \label{14} \\
&  R^{5(5D)}_5 =   \frac{ 8\pi G_5 }{3} \Big (-T^\mu_\mu + 2 T^5_5 \Big)   .  \label{15}
\end{align}
Multiplying equation \eqref{14} by $\frac{(1-n)W^n}{12}$ and equation \eqref{15} by $\frac{(n-4)W^n}{12}$ it can be obtained
\begin{align}
&\frac{(1-n)W^n}{12}R^{\mu(5D)}_\mu + \frac{(n-4)W^n}{12}R^{5(5D)}_5 =  \nonumber \\
& \frac{2}{3} \pi G_5 W^n  \Big [ T^\mu_\mu +  (2n-4)  T^5_5 \Big ].  \label{16}
\end{align}

By introducing $W=e^{-A(\phi)}$ in Req.\eqref{13} this transforms in
\begin{align}
-\frac{1}{r^2} \Big (A'e^{-nA}  \Big )'=&\frac{2\pi G_5}{3}e^{- nA}  \Big (T^\mu_\mu+(2n-4)T^5_5  \Big) \nonumber \\
                                                                &-\frac{1-n}{12}e^{(2-n)A}R^{(4D)}. \label{17}
\end{align}

Next, by noticing that the integration of the left side of equation \eqref{17} vanishes for compact internal spaces without boundary (for example RS), it is obtained
\begin{equation}
\int^\pi_{-\pi} d\phi e^{-nA}  \Big (T^\mu_\mu + (2n-4) T^5_5  \Big)=\frac{1-n}{8 \pi G_5}R^{(4D)} \int^\pi_{-\pi}d\phi  e^{(2-n)A}. \label{18}
\end{equation}

This is particular convenient if one has to consider matter fields constrained to the branes. For later convenience it is worth to explicitly mention the case $n=0$,
\begin{equation}\label{19}
\int^\pi_{-\pi} d\phi   \Big (T^\mu_\mu -4 T^5_5  \Big)=\frac{R^{(4D)}}{8 \pi G_5} \int^\pi_{-\pi}d\phi  e^{2A}.
\end{equation}

\section{ Calculation of the tensions of the two branes}
As mention above in this work it is considered only a modification of the tension brane which is replaced by (the conformal expansion on the FRW time) $T + \frac{t}{k} \frac{dT}{dt}$. Therefore, energy momentum tensor proposed including the two branes is given by
\begin{align}
T_{MN}=&- \frac{\Lambda_{5D}}{8 \pi G_5}  g_{MN}-  \Big  (T_1+ \frac{t}{k} \frac{dT_1}{dt} \Big )  h_{\mu\nu}^0 \delta^\mu_M \delta^\nu_N \delta(\phi) \nonumber \\
&-\Big (T_2+ \frac{t}{k} \frac{dT_2}{dt}  \Big )  h_{\mu\nu}^\pi \delta^\mu_M \delta^\nu_N \delta(\phi-\pi) + \tilde{T}_{MN}. \label{21}
\end{align}
Here $\tilde{T}_{MN}$ stands for the EM tensor of the rest of the matter fields, either constrained or not to the branes. In this toy model we have not considered equations of motion to be satisfied by the matter fields. This produces that in general some of the integration constants cannot be determined.

The $\phi-\phi$ component,
\begin{equation}
T^5_5= - \frac{\Lambda_{5D}}{8 \pi G_5} + \tilde{T}^5_5, \label{23}
\end{equation}
can be separated from the rest, giving rise to the partial trace
\begin{align}
T^\mu_\mu =& -4 \frac{\Lambda_{5D}}{8 \pi G_5} - 4  \Big  (T_1+ \frac{t}{k} \frac{dT_1}{dt}  \Big ) \delta(\phi)  \nonumber \\
&- 4  \Big  (T_2+ \frac{t}{k} \frac{dT_2}{dt}  \Big) \delta(\phi-\pi) + \tilde{T}^{\mu}_{\mu}. \label{22}
\end{align}

By replacing equations \eqref{22} and \eqref{23} at equation \eqref{19}:
\begin{align}
  T_1+ \frac{t}{k} \frac{dT_1}{dt}  + T_2+ \frac{t}{k} \frac{dT_2}{dt}   = U + V ,  \label{24A}
\end{align}
where
\begin{equation}
U=-\frac{R^{(4D)}}{32 \pi G_5} \int^\pi_{-\pi}d\phi  e^{2A} ,\label{25}
\end{equation}
and
\begin{equation}
V= \frac{1}{4} \int^\pi_{-\pi} d\phi   \Big (\tilde{T}^\mu_\mu -4 \tilde{T}^5_5  \Big).  \label{26A}
\end{equation}

The bulk vacuum case or when $\tilde{T}^\mu_\mu -4 \tilde{T}^5_5 $ vanishes the equation \eqref{24A} reduces to
\begin{align}
  T_1+ \frac{t}{k} \frac{dT_1}{dt}  + T_2+ \frac{t}{k} \frac{dT_2}{dt}   = U  ,  \label{24}
\end{align}

In next section it will be studied the cases where $T_1$, the tension of the brane that represents our universe, can be casted as
\begin{equation}
T_1=K x_c \Big ( 1 - \frac{a_{min}}{a(t)} \Big ),  \label{29}
\end{equation}
therefore when the brane corresponds to a E\"{o}tv\"{o}s branes.

It must be stressed that in this model $T_2$ is left partially undetermined by the equations of motions. The reason for this, as mentioned above, is that equations of motion of the five dimensional matter fields, and its boundary conditions, are not considered in this toy model.

\section{de-Sitter dS$_4$ scenario}

To begin with the discussion we will start with case $\tilde{T}^{MN} =0$. In this case we consider a flat transverse section in the brane, i.e., a metric of the form
\begin{equation}
ds^2 = e^{-2A(\phi)}(-dt^2 + a(t)^2 \mathbf{dx} \cdot \mathbf{dx}) + r^2 d\phi^2. \label{LE}
\end{equation}
The generalization to constant curvature transverse section is straightforward and does not provide new physical relevant information.

Remarkably this case contains as solution
\begin{equation}
a(t)=e^{H(t-t_0)} \label{31A}
\end{equation}
where $H$ is a constant and $t_0$ is the current age of our universe. $H$ corresponds to the standard Hubble constant (parameter). The tension of the de-Sitter brane is given by
\begin{equation}
T_1=U-\frac{C_3}{e^{H(t-t_0)}},  \label{32}
\end{equation}
which satisfies the E\"{o}tv\"{o}s law. By setting $U=K x_c$ (since $U$ is constant in dS$_4$ case)  and $C_3=a_{min} U= a_{min} K x_c$, the $T_1$ can be written as
\begin{equation}
T_1=K x_c \Big ( 1 - \frac{a_{min}}{ e^{H(t-t_0)}} \Big ).  \label{33}
\end{equation}
It must be noticed, since $U>0$, $T_1$ increases with $t$ (or with $a(t)$) yielding a brane which becomes more rigid as time evolves. On the other hand, from equation \eqref{24}, the strong brane tension,
\begin{equation}\label{35}
T_2=\frac{a_{min} K x_c}{e^{H(t-t_0)}},
\end{equation}
is positive. It is direct to observe that $T_2$ decreases with the time and therefore it become less rigid as time evolves.

It must be noticed that, even though this model is inspired by a generalization of a RS space, to immerse the branes into a negative cosmological space ($\Lambda_{5D} <0$) is not strictly necessary. This is due to the fact that $dS_4$ can be immersed into a $AdS_5$ as well as into a $dS_5$ or even into a five dimensional Minkowski space provided certain conditions are satisfied.

\subsection*{Constraints}

In the next table are shown the different warp factors, the corresponding $U$ and the constraints for $U>0$, such as $T_1$ increases with $a(t)$, at the scenarios $\Lambda_{5D} \to 0$ and $\Lambda_{5D} = \pm \frac{6}{l^2}$ respectively.

\begin{center}

 \begin{tabular}{ | c | c| c | c |}
    \hline
                                  & $e^{-2A(\phi)}$                                                                                                         & $U$                                                                                &constraint \\ \hline
     $ \Lambda_{5D} \to 0$   &  $ \big (|\phi| - K_1 \big )^2    \frac{H^2 r^2}{4}$                                                    &   $-\frac{3}{4r^2K_1 G_5 (K_1-\pi)}$                          & $K_1<\pi$      \\ \hline
     $ \Lambda_{5D} < 0$     &  $ \big ( \frac{Hl}{2} \big)^2 \sinh ^2 \big (K_1-\frac{r}{l}|\phi| \big) $                          & $ \frac{3}{4}\frac{\coth(K_1)-\coth(K_1-\frac{\pi r}{l})}{l r \pi G_5 } $     & $ K_1<\frac{\pi r}{l} $   \\    \hline
    $ \Lambda_{5D} > 0$      & $\big ( \frac{Hl}{2} \big)^2 \sin ^2 \big (K_1 \pm\frac{r}{l}|\phi| \big)$                                                           & $\frac{3}{4}\frac{\cot(K_1)-\cot(K_1 \pm \frac{\pi r}{l})}{l r \pi G_5 }$                                                                                & * \\ \hline
  \end{tabular}
\end{center}
where $K_1$ is a constant, and the $*$ at the table means that for different values of $K_1$ , the integral $U$ could be positive, due to the periodicity of trigonometric functions.

\section{Radiation domination}

The next simplest solution can be obtained by imposing the vanishing of the four dimensional Ricci scalar, $R^{(4D)}=0$. See Eq.(19). For this let us consider a general energy momentum tensor, only constrained by the symmetries of the space,
\begin{align}
&\tilde{T}^C_N = \mbox{diag} \Big  (-\rho(t,\phi), p(t,\phi), p(t,\phi), p(t,\phi), p_5(t,\phi)  \Big ). \label{35A}
\end{align}

Firstly, to have a solution with the line element \eqref{LE} is necessary to be restricted to the case $\Lambda_{5D} <0$. It is direct to prove, as for standard cosmology, that the direct solution for $R^{(4D)}=0$ is $a(t) = (t/t_0)^{1/2}$ where $t_0$ is the current cosmological time. However, and unlike the standard cosmological case, in this case this restricts a set of functions of $\phi$ and $t$. The Einstein equations determine that $p_5(t,\phi)=0$ and
\begin{equation}\label{39A}
A(\phi)=\frac{r}{l}|\phi|,
\end{equation}
where the negative five dimensional cosmological constant has been fixed as $\Lambda_{5D}=-\frac{6}{l^2}$. Finally, the Einstein equations also determine that
\[
\rho(t,\phi)=\frac{3}{4} \frac{e^{\frac{2r}{l}|\phi|}}{t^2} \textrm{ and } p(t,\phi)= \frac{1}{4} \frac{e^{\frac{2r}{l}|\phi|}}{t^2}.
\]

As the trace of $\tilde{T}^{MN}$ vanishes as well as its partial trace (four dimensions), i.e., $(\tilde{T}^M_{M} = \tilde{T}^{\mu}_{\mu} = 0)$ this energy momentum tensor seems to correspond to a fluid that is a generalization of an electromagnetic field. Notice that the energy momentum tensor, unlike its trace, does not vanish outside of the branes but it extends into space between both branes. One can argued, by observing the components, that this energy momentum tensor can be generated by the combination of a five dimensional pressureless fluid and a fluid of string-like objets with endpoints at both branes and a two form field of the form $B(x^{\mu},\phi) = A(x^{\mu},\phi)_{\mu} dx^{\mu} \wedge d\phi$ which couples the string-like fluid. The pullback of $B$ into the branes, $A$, can be interpreted as the EM gauge potential.

\subsection*{Tension of both branes}
As $V=0$ and $\tilde{T}^\mu_\mu -4 \tilde{T}^5_5 $ vanishes, then Eq.(\ref{24}) implies
\begin{equation}
T_1+T_2= - \frac{C}{t^k}.
\end{equation}
where $C$ and $k$ are arbitrary. Now, in order to have a consistent set of equations and fix $T_1$ into the form of an E\"otv\"os brane, is necessary that $k=1/2$. This implies that the tensions are respectively,
\begin{align}
T_1=& H_1- \frac{C_4}{t^{1/2}} \label{C1} \\
T_2=&-H_1+ \frac{C_5}{t^{1/2}} \label{C2},
\end{align}
where $C=C_4-C_5$ and $H_1$ are arbitrary constants. To cast $T_1$ into the form of an E\"otv\"os brane form in Eq.(\ref{3}) it is  necessary to fix the constant as $H_1=K x_c$ and $C_4=a_{min}t_0^{1/2} H_1$.

As mentioned above $T_2$ is left partially undetermined. The simplest solution is $T_2=-K x_c$ ($C_5=0$) leaving a negative, but constat, tension brane.

\section{Matter dominated era}

In the model proposed, for line element \eqref{LE},  is also possible to obtain a solution with $A(\phi)$ similar to equation \eqref{39A} and $a(t)=(t/t_0)^{2/3}$  which corresponds, from the point of view of the universe brane, to a matter dominated era in a standard cosmological model. For this to happen $p(t,\phi)=0$ must be satisfied as well. In addition
\[
\rho(t,\phi)=\frac{4}{3} \frac{e^{2A(\phi)} }{t^2}.
\]
Finally, the rest of the Einstein equations implies that
\[
p_5(t,\phi) = -\dfrac{2}{3t^2} e^{2A(\phi)}
\]
which implies that although there is no pressure along the brane directions it must exist along the fifth dimension for an equilibrium to exist. This can be modeled by a pressureless (particle) fluid combined with non-fundamental string like fluid with endpoints at the branes.

\subsection*{Tensions of both branes}
In this case Ricci Scalar
\[R^{(4D)}=\dfrac{4}{3t^2}.
\]
and from \eqref{25} and \eqref{26A} one can identify that
\[
U= \frac{l}{r t^{2}} \left(\dfrac{1-\exp(2\pi \frac{r}{l})}{24 \pi G_5 }\right) \textrm{ and } V = \frac{l}{3r t^2}\left(\exp\left(2\pi \frac{r}{l}\right)-1\right).
\]
Now, by defining $U+V = \dfrac{m}{t^2}$, where $G_5$ is dimensionless, then it can be shown that $T_1$, the tension of our brane universe, can be shaped up into the form of the tension of an E\"{o}tv\"{o}s brane,
\[
T_1(t) =K x_c\left(1-\dfrac{t_0^{2/3}}{t^{2/3}}\right).
\]

On the other hand, from equation \eqref{24A}, the tension of strong brane is given by
\[
T_2(t)= - K x_c + \dfrac{K x_c t_0^{2/3}}{t^{2/3}}- \dfrac{m}{2t^2}.
\]
This tension can be considered a generalization of the tension of an E\"{o}tv\"{o}s brane, with a negative sign.

\section{Conclusions}

Using the modification of energy momentum tensor of equation \eqref{21}, we have shown that it is possible the study the branes with temporally variable tension using Brane World Sum Rules. Specifically we have studied branes with variable tensions that resemble E\"{o}tv\"{o}s branes.

In particular we have shown an alternative to obtain the scale factors similar to the obtain from FRW for the component individual of energy cases (matter, radiation and cosmological constant cases).

Also we have shown three cases where the same scale factor is obtain from geometry $4D$ and from tension of our brane universe. This cases are : FLRW $4D$ brane with radiacion and matter domination on a $5D$ space, and dS$_4$ (\emph{i.e} de Sitter universe with exponential scale factor) on a (A)dS$_5$.

Finally it has been shown the different values for the warp factor, the integral $U$, and constraints for that $U>0$ at the cases $\Lambda_{5D} \to 0$ and the (A)dS$_5$ cases.

\begin{acknowledgements}
Milko Estrada acknowledges Julio Hoff da Silva for useful conversations. This work was partially founded by FONDECYT Regular Grants 1151107, 1131075, 1140296 and UNAB DI-735-15/R.

\end{acknowledgements}

\end{document}